\documentclass[twocolumn,superscriptaddress, floatfix, showpacs, aps, prb, 10pt]{revtex4-2}
\usepackage{physics}
\usepackage{color}
\usepackage{graphicx}
\usepackage{dcolumn} 
\usepackage{bm} 
\usepackage[normalem]{ulem}
\usepackage{breqn}

\usepackage{hyperref}
 \hypersetup{
 colorlinks = true, 
 urlcolor = blue, 
 linkcolor = blue, 
 citecolor = red 
 }

\newcommand{\INFN}{INFN - Sezione collegata di Salerno, Complesso Univ. Monte S. Angelo, I-80126 Napoli, Italy}

\newcommand{\UNISA}{Physics Department ``E.R. Caianiello'', Universit\`a degli Studi di Salerno, Via Giovanni Paolo II, 132, I-84084 Fisciano (Sa), Italy}

\newcommand{\TUWIEN}{Institute of Solid State Physics, TU Wien, 1040 Vienna, Austria}

\begin{document}

\title{Engineering superconductivity on the surface of Weyl
semimetals}

\author{Riccardo Vocaturo}
\email{riccardo.vocaturo@tuwien.ac.at}
\affiliation{\TUWIEN}

\author{Mattia Trama}
\email{mtrama@unisa.it}
\affiliation{\UNISA}
\affiliation{\INFN}

\date{\today}

\begin{abstract}

	Ten years after the experimental discovery of Weyl semimetals,
	theoretical and experimental work has pointed to the possibility of
	realizing surface-only superconductivity at relatively high temperatures
	in these materials. A consensus is developing that this unusual form of
	superconductivity is {supported} by surface electronic states unique to
	Weyl semimetals, known as Fermi arcs.
	In this work, we show that the topological protection of these exotic states 
    can be exploited to engineer
	high critical temperatures. {Motivated by reports of van Hove singularities on the surface of PtBi$_2$---the first Weyl semimetal in which surface superconductivity was unequivocally linked to its Fermi arcs---we demonstrate how such singularities can be engineered through the deposition of a suitable additional layer. We find that the critical temperature becomes extremely sensitive to the exact value of the chemical potential, and values as large as 20 times the bulk $T_c$ can be achieved.}
	More generally, our results demonstrate how topological protection
	can be exploited to manipulate surface electronic
	states, thereby opening experimentally accessible routes toward
	engineering high-temperature two-dimensional superconductivity
	and other exotic phases.
\end{abstract}

\keywords{Weyl semimetals, surface superconductivity, surface engineering}
\maketitle


{\textit{Introduction.}} ---
Topological materials are systems hosting robust surface states, whose presence is dictated by the topology of their bulk electronic band structure~\cite{Review1,Review2}.
Specifically, Weyl semimetals are a type of topological matter characterized by unique kind of surface states known as Fermi arcs~\cite{review_weyl,bo2015crossed}. 
Unlike {trivial} surface states, which arise from termination-dependent effects such as surface reconstruction, impurities, or dangling bonds, the existence of Fermi arcs is guaranteed by the topology of the bulk Hamiltonian. As a consequence, weak local perturbations cannot remove these states.

\begin{figure}[ht]
	\centering
	\includegraphics[width=0.47\textwidth]{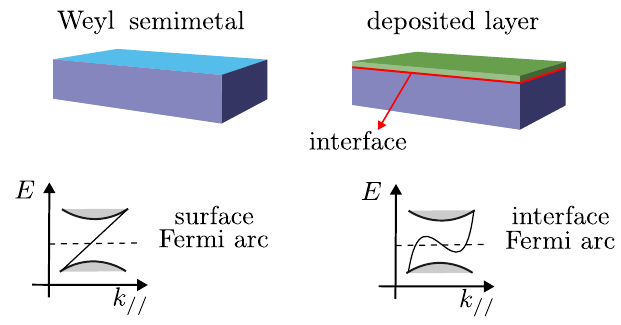}
	\caption{(Left) Weyl semimetals host topologically protected surface states
		(Fermi arcs). The surface is highlighted in light blue and a
		representative surface band structure showing a Fermi arc crossing
		the Fermi level is shown below.  (Right) Schematic effect of
        depositing an atomic layer on the surface: the Fermi arc hybridizes with the overlayer, changing its dispersion.}
	\label{fig:idea}
\end{figure}

Remarkably, this robustness can be practically exploited through surface or interface engineering.
For example, consider depositing one or more atomic
layers of a different material on top of a Weyl semimetal substrate. 
Since a finite overlayer cannot alter the topology of the substrate,
the surface Fermi arcs cannot be removed. Instead, they will localize at the interface between the Weyl semimetal and the overlayer~\cite{Armitrage2018}.
However, the interface can induce a redistribution of spectral weight and modify the dispersion of the Fermi arcs, enabling controlled tuning of the surface (or interface) band structure.
We present a simple sketch of the set up in Fig.~\ref{fig:idea}.

So far, this effect has been theoretically studied only for interfaces between different Weyl semimetals~\cite{Dwivedi2018,Ishida2018}, but its applicability is rather general.
In this work we investigate whether
a robust quasi-two-dimensional, superconducting phase can be engineered at the interface between a time-reversal invariant Weyl semimetal and a deposited film. We show that, upon a suitable choice of the deposited film, Fermi arcs can be made unstable towards a superconducting order with significant critical temperature ($T_c$).

Our study is further motivated by recent reports of 
surface-only superconductivity at relatively high temperatures in the topological semimetals PtBi$_2$ and
ZrAs$_2$~\cite{Kuibarov2024,Schimmel2024,Hossain2025}.
Specifically, many experiments carried out on PtBi$_2$ samples confirm that the unusual
superconducting state is indeed mediated by the topological Fermi arcs~\cite{Kuibarov2024,Kuibarov2025,Schimmel2024,besproswanny2025,qpi}. For the same system,
scanning tunneling microscopy also shows that the superconducting gap is highly
inhomogeneous, reaching values as large as 20 meV~\cite{Schimmel2024,Schimmel2025}. 

{Interestingly, despite a report of Berezinskii-Kosterlitz-Thouless~\cite{veyrat2023berezinskii} transition in relative thick samples of PtBi$_2$, there is no evidence of such transition when superconductivity is localized on the surface. New experimental evidence on PtBi$_2$ supports a Bardeen–Cooper–Schrieffer (BCS)-like behavior~\cite{moreno2026robust}}.
A BCS-based argument suggests that
Fermi arcs enhance the local density of states (DOS) at the surface and thus
increase $T_c$. However, it is hard to account for the large variations across different samples observed in experiments, especially since all samples are reported to have the same nominal purity~\cite{Cristalli}.

Interestingly, \textit{ab initio} density-functional-theory (DFT) calculations
corroborated by angle-resolved photoemission spectroscopy (ARPES) reveal the presence of surface van Hove singularities (VHSs) in both PtBi$_2$ and ZrAs$_2$~\cite{Kuibarov2025,Hossain2025}. In
PtBi$_2$ the singularity lies only 6~meV below the
Fermi level.

We recall that VHSs are divergences in the single-particle DOS.
Thus, even small amounts of impurities can shift
the surface chemical potential closer to such singularities, boosting $T_c$~\cite{Hirsch1986}. 
This mechanism also naturally accounts for the large fluctuations observed in different samples. In fact, a numerical study showed that $T_c$ is extremely sensitive to the position of the chemical potential when close to a VHS~\cite{Mahan1993}.

Motivated by these findings, we propose to engineer VHSs on the surface of Weyl
semimetals via layer deposition or, more generally, interface engineering.  We
consider an effective model for PtBi$_2$-like semimetals and we show how surface
VHSs originate from long-range hoppings. Importantly, we show that long-range
hoppings need only be engineered on the surface layer, rather than through the
entire material, and that they naturally lead to enhanced surface
superconductivity.

\begin{figure}[t]
	\centering
	\includegraphics[width=0.47\textwidth]{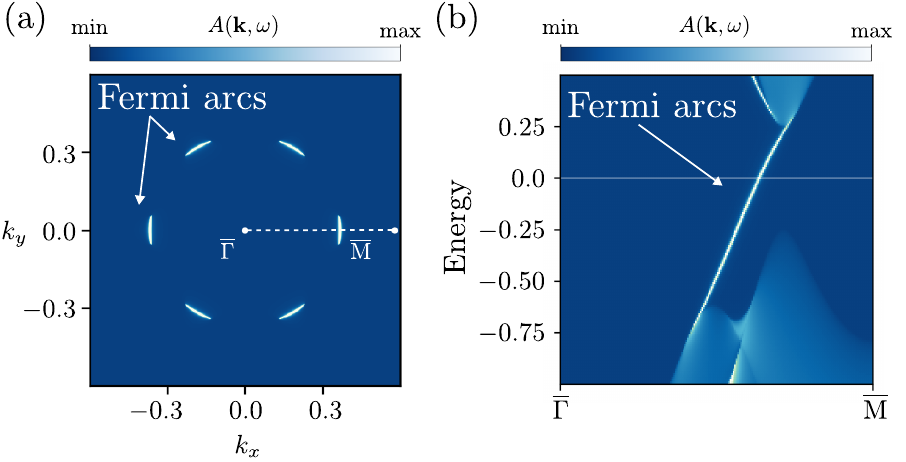}
	\caption{Normalized surface spectral function of the model from Ref.~\cite{Vocaturo2024}. (a) Constant energy plot at the Fermi level. (b) Energy distribution curves along the $\overline{\Gamma \hbox{M}}$ line highlighted in (a). }
	\label{fig:tb_sketch}
\end{figure}

{\textit{Fermi arcs and van Hove singularities}.} ---
We start from the four-band tight-binding model introduced in Ref.~\cite{Vocaturo2024} as a minimal description of the time-reversal invariant Weyl semimetal PtBi$_2$. {The model sits on a triangular lattice, having three mirror planes along the in-plane lattice vectors. It has point group $C_{3v}$ and belongs to space group $P31m$. The Weyl points are generated by breaking inversion-symmetry via a local on-site hopping between the two orbitals.} For completeness, we report the bulk Hamiltonian in the Appendix~\ref{app:ham}.
We fix the nearest-neighbor (NN) hopping to $t = 1$, so that all energies are expressed in units of $t$. The model is defined on a trigonal lattice and hosts six symmetry-related Fermi arcs on the surface Brillouin zone (BZ). We can resolve them by plotting the surface spectral function
(Fig.~\ref{fig:tb_sketch}).

The localization of Fermi arcs on the surface is not uniform in reciprocal space. It is known that Fermi arcs progressively become bulk-like close to their end points, while achieving maximal localization further away from them~\cite{PenetrazioneArchi}. 
For now, we treat the arcs as fully two-dimensional; we will address this later.
Under this assumption, a VHS can be obtained by introducing a saddle point in the surface energy-momentum dispersion~\cite{Ziman}.

\begin{figure*}
\centering
	\includegraphics[width=0.9\textwidth]{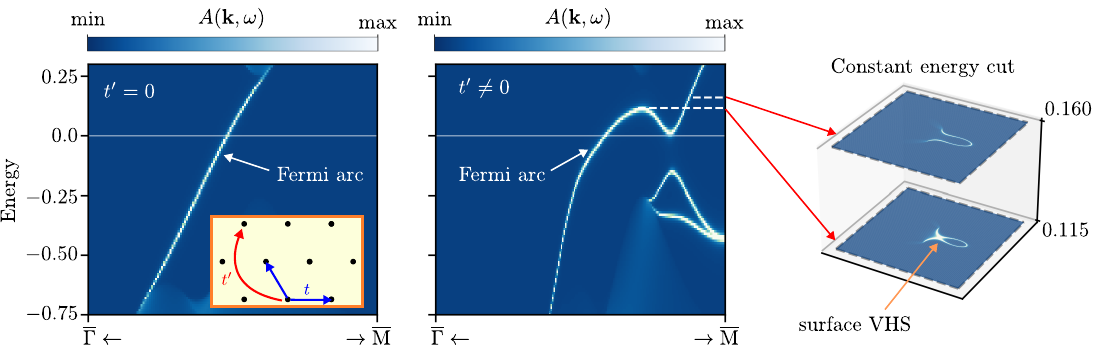} 
	\caption{Left:
	surface normalized spectral function for the tight-binding model with and without $t'$ (shown in inset). Middle: surface normalized spectral function for $t' = 0.6$. Right: Zoom-in showing constant-energy maps taken at the energies marked by the white dashed lines. The VHS sits roughly at $E \sim 0.115$.} 
	\label{fig:arcs_t3}
\end{figure*}

In this model, mirror and time-reversal symmetry force the arcs to be symmetric with respect to the $k_y = 0$ line~\cite{Vocaturo2024}. In particular, we find that for any $k_x \neq 0$, the $y$-component of the band velocity has a local minimum along $(k_x,\delta k_y)$ for small $\delta k_y$. The Fermi arcs' dispersion, however, increases monotonically along $k_x$, as shown in Fig.~\ref{fig:tb_sketch}(b). 
It follows that the simplest way to generate a saddle point is to introduce
third-nearest-neighbor hoppings such as the one depicted in the inset of Fig.~\ref{fig:arcs_t3}. These hoppings, denoted as $t'$, yield a contribution $-2t'\cos(2k_x)$ that modulates the otherwise monotonic dispersion along
$k_x$, and allow the VHS to appear~\footnote{All symmetry-related terms are also included by construction so six symmetry-related VHSs appear.}.
Since fine tuning hoppings in the bulk of a crystal is generally challenging, we show that this mechanism is effective even when $t'$ is introduced only at the surface level. 
{We restricted ourselves to the addition of $t'$ because the symmetries of the lattice  forces next-nearest-neighbor to give a monotonic correction along $\overline{\Gamma \mbox{M}}$. Their effect is therefore mainly to change the values of $t'$ for which the singularity is observed.}

To demonstrate this, we plot the surface spectral function of our model modified by the addition of $t'$ to the top layer. The procedure for obtaining the new spectral function is detailed in Ref.~\cite{VocaturoPhD}. 
The result is plotted
in Fig.~\ref{fig:arcs_t3} alongside the spectral function from Fig.~\ref{fig:tb_sketch}(b) for comparison. The monotonic dispersion along $\overline{\Gamma \mbox{M}}$ is broken, and a VHS emerges around $E = 0.115$ (orange arrow). In the Appendix~\ref{appendix:vhs} we provide a 3D representation of the energy dispersion to highlight the saddle point.
The emergence of the VHS is accompanied by the distortion of the arc into a horseshoe-shaped contour. This behavior is consistent with that observed in DFT calculations for PtBi$_2$~\cite{Kuibarov2025}, suggesting that the mechanism proposed here captures the underlying physics.

We now return to the discussion of the effective dimensionality of these surface
states because the dimensionality ultimately defines the nature of the singular
point (logarithmic divergence or kink in the DOS)~\cite{Ziman}. The VHS is here
located on the $\overline{\Gamma \mbox{M}}$ line, which passes through the
middle of the arc. The points belonging to this line turn out to be where the
arcs are (almost) maximally localized~\cite{Vocaturo2024,PenetrazioneArchi}. 
This observation justifies \textit{a posteriori} treating the VHS as effectively two-dimensional, thus expecting a logarithmic divergence in the surface DOS. However, numerical confirmation of this statement is challenging since the arcs occupy a tiny portion of the surface BZ and very fine sampling is needed to reproduce the singular behavior.

In Fig.~\ref{fig:arcs_t3}, additional spectral weight unrelated to the Fermi arcs is also seen at higher energies. This is expected since
the addition of $t'$ on the surface renormalizes the dispersion of all states belonging to the top-most layer (topological ones as well as trivial bulk resonances). These states, however, will not contribute to superconductivity.

We conclude this section with a final remark about the magnitude of the
hoppings. 
Our results were obtained setting $t'=0.6$.
While this large value may be unrealistic in practice, the minimum value of $t'$
that induces the VHS is close to $t'=0.5$, and could be engineered to be even
 {upon different choice of parameters}.  The striking similarity with the Fermi arcs of PtBi$_2$ obtained once
$t'$ is included is an encouraging result.


{\textit{Superconductivity.}} ---
We quantify the impact of the VHS on the onset of superconductivity.
Since an exact two-dimensional Hamiltonian for the surface arcs cannot be derived, one should in principle solve the Bogoliubov–de Gennes {(BdG)} equations for a sufficiently large system to properly capture their contribution.
However, such simulations are computationally demanding.
For this reason, we adopt an alternative variational approach based on free-energy minimization, developed in Refs.~\cite{Trama2025,trama2026inhomogeneous}.
We assume an onsite, intra-orbital attractive interaction,
$H_{\mathrm{int}}=-U\sum_{\mathbf r,z,\alpha}
n_{\mathbf r z\alpha\uparrow}n_{\mathbf r z\alpha\downarrow}$,
with $U>0$. We restrict the mean-field solution to an
orbital-independent spin-singlet order parameter.
{The resulting multi-orbital mean-field Hamiltonian
for a semi-infinite slab is}
\begin{equation}
\begin{split}
H=H_0
&-U\sum_{\mathbf{k}\in\frac{1}{2}\mathrm{BZ}}
\sum_{z,\alpha}
\left(
\Delta_z\,
c^\dagger_{\mathbf{k}z\alpha\downarrow}
c^\dagger_{-\mathbf{k}z\alpha\uparrow}
+\mathrm{h.c.}
\right)
\\
&+ U\sum_{z,\alpha}|\Delta_z|^2,
\end{split}
\label{eq:mean_field_super}
\end{equation}
with $\mathbf{k}=(k_x,k_y)$, and $H_0$ being the single particle Hamiltonian.
In Eq.~\eqref{eq:mean_field_super} the system is assumed homogeneous across the $xy$-planes, while non-homogeneity is allowed only along $z$ (the direction perpendicular to the surface): at $z=0$, we include a NNN hopping term $t'$ in the single-particle Hamiltonian, whereas for $z \geq 1$ one has $t'=0$. The subscript $\alpha$ refers to the orbital degrees of freedom and $H_0$ is the tight-binding model introduced before. 
The superconducting order parameter is defined as
\begin{equation}
\Delta_{z\alpha}
=\langle c_{\mathbf r z\alpha\uparrow}
          c_{\mathbf r z\alpha\downarrow}\rangle
\equiv\Delta_z .
\label{eq:def_gap}
\end{equation}
{which is what we expect after mean-field decoupling of a local interaction belonging to the trivial $A_1$ irreducible representation of the point group (here $C_{3v}$). This is the simplest choice of pairing allowed by time-reversal symmetry. Nonetheless, Eq.~\eqref{eq:def_gap} does not exclude a mixture of singlet and triplet channels when projecting onto the Fermi arc states~\cite{Vocaturo2024}. 
{In fact, in the spin-space representation of the projected Fermi-arc gap introduced in the Appendix~\ref{app:spin_gap_symmetry}, the spin texture originated by spin-orbit-coupling and broken-inversion symmetry gives rise to a $p$-wave-like contribution.} 
Due to the exponential localization of the surface states, we consider solutions of the form
\begin{equation} 
\Delta_z = \Delta_0 + \Delta_1 e^{-z/\lambda}    
\label{eq:pairing}
\end{equation}
Eq.~\eqref{eq:pairing} 
captures both bulk superconductivity ($\Delta_0$) and the additional surface contribution ($\Delta_1$).
The exponent quantifies the penetration of Cooper pairs into the bulk and $\lambda$ is the associated characteristic length and is determined self-consistently. {This choice of functional form is discussed in great lengths in Ref.~\cite{Trama2025} and has been validated against numerical results for the non-homogeneous BdG equations on a finite system.~\cite{Trama2025,trama2026inhomogeneous,Bai2025}.}

Within this ansatz, the superconducting free-energy functional simplifies to
\begin{equation}
    F[\Delta_z] = F_B[\Delta_0] + F_S[\Delta_0,\Delta_1,\lambda].
    \label{eq:free_energy}
\end{equation}
$F_B$ accounts for conventional bulk superconductivity and, for a homogeneous system, favors the BCS ground state with a constant gap $U\Delta_0$. Surface superconductivity is described by $F_S$. 
$F_B$ scales with the volume of the system whereas $F_S$ does not, and becomes negligible in the thermodynamic limit.
This behavior allows for a two-step self-consistent solution: (i) we compute $U$ and $\Delta_0$
fixing $U\Delta_0=10^{-4}$ at $T=0$ via self-consistent bulk calculation; 
(ii) we use the obtained $\Delta_0$ and minimize $F_S$ with respect to the remaining parameters. Additional details on the method are deferred to the Appendix~\ref{appendix:sc}.
We stress that, despite having two order parameters $\Delta_0$ and $\Delta_1$, we fix only one electron-electron interaction constant $U$. In this way, Eq.~\eqref{eq:free_energy} treats bulk and surface on equal footing and does not introduce additional biases for the mechanism of superconductivity.

In Fig.~\ref{fig:scf} we show in blue the total gap collected at the surface ($z=0$) as a function of the chemical potential $\mu$. A spike in the energy of the VHS is observed; it corresponds to a maximum value more than 30 times larger than that in the bulk. Interestingly, if we set $t = 0.75$~eV to match the experimental bulk critical temperature $T_c^{\text{bulk}} $ in PtBi$_2$, we extrapolate a surface critical temperature $T_c^{\text{surf}} \sim 13$~K, in agreement with STM and ARPES data~\cite{Schimmel2024,Kuibarov2024}. Further details on the temperature dependence order parameter can be found in the Appendix~\ref{appendix:sc}.

\begin{figure}[t]
	\centering

    \includegraphics[width=0.47\textwidth]{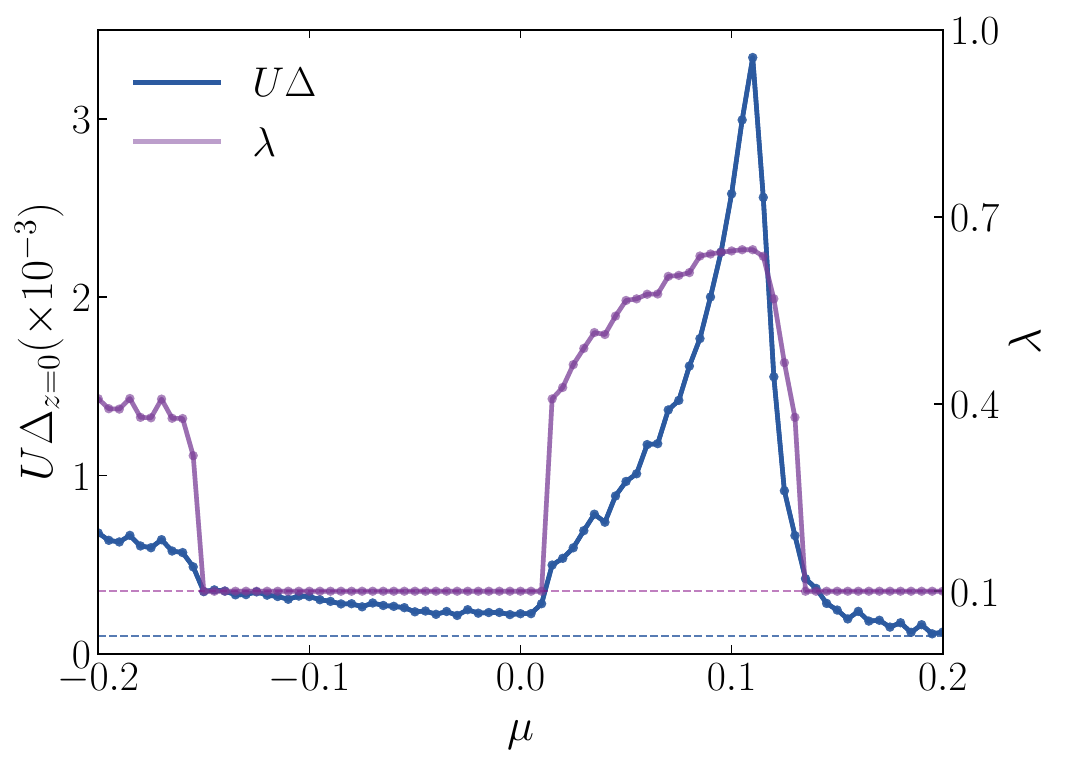}
	\caption{In blue: full self-consistent gap $U\Delta$ at zero temperature as a function of the chemical potential. There is a significant spike corresponding to the energy of the VHS (see Fig.~\ref{fig:arcs_t3}).
    In purple: penetration depth ($\lambda$ in Eq.\eqref{eq:pairing}) as a function of the chemical potential. The blue dashed line indicates the bulk gap limit $U\Delta_0 = 10^{-4}$, while the purple dashed line marks the minimum computed value of $\lambda$. {For computational reasons we resolved $\lambda$ only down to $0.1$.}}
	\label{fig:scf}
\end{figure}

It is also instructive to point out that around $\mu=-0.2$ there are additional surface states with comparable superconducting penetration length $\lambda$ (in orders of magnitude). As discussed above, these states originate from the hybridization induced by surface hoppings ($t'$), which acts as a local impurity. 
However, the corresponding $U\Delta$ is vanishingly small. This supports the conclusion that the enhancement of superconductivity is linked to the VHS and not merely to the presence of any (trivial) surface state.

{\textit{Surface engineering.}} ---
Having established a connection between long-range hoppings, VHSs, and surface superconductivity, we look for a practical procedure to engineer the required hoppings ($t'$ in our model) at a generic surface.
To this end, we identify film-deposition as a powerful tool. This is because, 
due to the presence of the interface, we provide an alternative path for surface electrons to 
effectively hop between next-nearest-neighbors (NNN) by going through the overlayer. 

To show this,
let us consider the setup illustrated in Fig.~\ref{fig:top_layer}: the substrate
consists of two orbitals per unit cell, labeled $A$ and $B$, forming a bipartite
lattice. An additional overlayer composed of a different atomic species $C$ is
deposited on top. 
In this example, the $AB$-system 
represents
the Weyl semimetal substrate. 
For the purpose of this section we simply describe it as a one-dimensional chain of
alternating $A$ and $B$ atoms connected only through
NN hopping $t$ between $A$ and $B$. 

The
Hamiltonian for a supercell of 2 units reads:
\begin{equation} 
    H_{AB}(k) = \begin{pmatrix} \varepsilon_{A} & t & 0 &
    te^{-i2ka} \\ t & \varepsilon_{B} & t & 0 \\ 0 & t & \varepsilon_{A} & t \\
    te^{i2ka} & 0 & t & \varepsilon_{B} 
    \end{pmatrix}.  
    \label{eq:1d_chain}
\end{equation}
This choice of basis makes it clear that there are no long-range hoppings because of the null entries between the $A$ site in cell $i$ and $A$ site in cell $i+1$.
\begin{figure}[ht]
	\centering
	\includegraphics[width=0.47\textwidth]{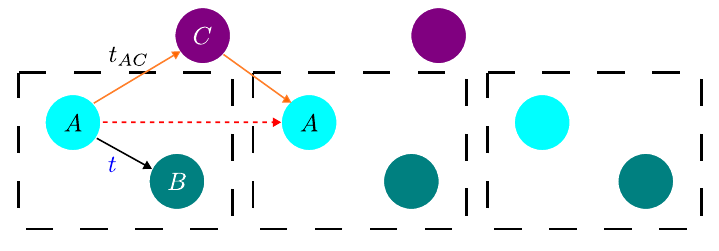}
	\caption{Schematic representation of the proposed set up for an interface. The substrate is comprised of 2 atoms per unit cell, here represented
as an alternating 1D chain with NN hoppings $t$
between A and B. On top a new layer ($C$) with
strong chemical affinity for A added.
An effective long-range hopping between adjacent unit cells $t_{AA}$ develops (dashed line).}
	\label{fig:top_layer}
\end{figure}

We now examine the effect of adding the overlayer ($C$ in Fig.~\ref{fig:top_layer}). If the atomic species $C$ has
strong chemical affinity with $A$, but negligible coupling to $B$, a sizable
hybridization $t_{AC}$ develops. Electrons can then hop from $A$ to $C$ and back to a neighboring $A$ site, thereby generating an effective channel that bypasses the intermediate $B$ site.
For simplicity, we assume hoppings between $B$ and $C$ to be zero so that the coupling matrix between $AB$ and $C$ is given by:
\begin{equation} 
    V = \begin{pmatrix} 0 & t_{AC} & 0 & 0 \\ t_{AC} & 0 &
	t_{AC} & 0 \\ 0 & t_{AC} & 0 & 0 \\ 0 & 0 & 0 & 0 
    \end{pmatrix}.
\end{equation}

Integrating out the overlayer yields the effective interface Green's function: 
\begin{equation} 
    G_{AB}^{\text{eff}}(k,\omega) = \left[ \omega - H_{AB}(k) - V G_C( k, \omega)
    V^\dagger \right]^{-1}, 
    \label{eq:surf_g}
\end{equation}
where $VG_C(k, \omega)V^{\dag}$ is commonly referred to as \textit{hybridization function} and renormalizes the original dispersion described by $H_{AB}(k)$. To treat the problem analytically, 
we consider $C$ to be a collection of localized atomic orbitals with negligible $C-C$ overlap.
With our current notation we have:
\begin{equation} 
    H_C = \mathrm{diag}(0,\varepsilon_C,0,\varepsilon_C).
\end{equation}
Since we did not assume strong-coupling in our previous analysis of superconductivity, we restrict ourselves to a small energy window close to the Fermi level so that $\omega \ll \varepsilon_C$. In this case, we take the static limit ($\omega \rightarrow 0$) and invert Eq.~\eqref{eq:surf_g} to write the following effective Hamiltonian: 
\begin{equation} 
    H^{\mathrm{eff}}_{AB}(k) = \begin{pmatrix}
	\varepsilon_{A} + t' & t & t' & te^{-i2ka} \\ t & \varepsilon_{B} & t &
    0 \\ t' & t & \varepsilon_{A} + t' & t \\ te^{i2ka} & 0 & t & \varepsilon_{B}
    \end{pmatrix}  
\end{equation}
with $t' = -\frac{t_{AC}^2}{\varepsilon_C}$. Comparison with Eq.~\eqref{eq:1d_chain}
shows that matrix elements corresponding to
longer-range hoppings between $A_i$ and $A_{i+1}$ are now finite.
This construction demonstrates that selective hybridization with an
appropriately chosen overlayer provides a controlled route to engineer effective
surface longer-range hopping in the substrate. 

This simple argument also introduces a natural way to quantify the effect of hybridization, which can serve as a starting point to implement high-throughput search algorithms. For example, for an insulating overlayer, $\varepsilon_C$ can be replaced by the insulating gap $\Delta E$. Then, once $t_{AC}$ and $\Delta E$ are extracted from a DFT calculation, the ratio $t'/t$ can be estimated. Larger values of this ratio indicate a higher likelihood of observing a VHS. Knowledge of the trend of $t'/t$ for different interfaces then enables the implementation of screening schemes to guide experiments towards the realization of such systems.

{\textit{Summary}.} ---
We investigated the interplay between long-range hoppings and the appearance of VHS on the surface of PtBi$_2$-like topological semimetals. The combination of time-reversal and crystal symmetries of this lattice~\cite{Vocaturo2024} establishes a simple relation between third NN hoppings and the energy profile of the topological surface states, i.e. the Fermi arcs. In particular, we show that VHS can be fine-tuned by the relative strength of such hoppings with respect to the NN ones.  

We studied the superconducting properties of these VHSs within a mean field variational approach for a general $s$-wave pairing. 
Among all contributions coming from the top layer  we clearly observe a significant enhancement of surface superconductivity linked to the presence of the singularities.
The onset of surface superconductivity as a function of the chemical potential is found to be sharply asymmetric with respect to the energy of the VHS. 
This behavior provides a natural explanation for (i) the increase of surface $T_c$ observed in PtBi$_2$ from a bulk value of $0.6$~K up to $T_c^{\text{surf}} \sim 13$~K;  (ii) the difficulty in consistently observing superconducting gaps in PtBi$_2$~\cite{Kuibarov2025_2,Schimmel2024,oleary2025topography}; and (iii) the significant variation of the gap size across samples.
Such a strong sensitivity to the chemical potential has not been reported in previous theoretical studies~\cite{Bai2025,Boh2025,Nomani2023}.
However, those works did not consider the presence surface VHSs, which naturally accounts for the discrepancy.
We point out that our analysis is easily extended to order parameters with
higher angular momentum. This might be necessary to quantitatively account for
new experimental evidence regarding the symmetries of the superconducting gap in
PtBi$_2$~\cite{J-iwave,maeland2025mechanism}.
{We believe the mechanism discussed here is also relevant to other materials. For example, DFT calculations have shown that B20 chiral Weyl semimetals such as RhSi, CoSi, and BeAu host large Fermi arcs with different types of surface VHSs~\cite{Sanchez2023,Vocaturo2026}.}

Finally, we have presented a theoretical argument to show that engineering surface VHSs via interface design is, in principle, feasible. 
Although the experimental realization of such interfaces remains challenging, we propose a simple criterion to assess the potential of candidate materials based on relatively inexpensive DFT calculations. 
This approach enables pre-screening of promising systems through high-throughput or machine-learning-assisted searches of materials databases.

{\textit{Acknowledgments}.} ---
We thank V. Konye, I. C. Fulga, J. van den Brink for useful discussions. M.T thanks R. Citro, C. Guarcello and D. F. G. Fiorillo for useful discussion and support.
M.T. acknowledges funding from acknowledges support from the PNRR MUR Project No. PE0000023-NQSTI (SPUNTO) and computational resources from CINECA award under the ISCRA C initiative (project DISKTO HP10CCMDC7). RV acknowledges funding from the FWF project DOI 10.55776/P36213.

\newpage
\appendix
\onecolumngrid
\section{Effective model for the bulk}
\label{app:ham}

The effective bulk Hamiltonian used in the main text is the one used in Ref.~\cite{Vocaturo2024}, which we report here for completeness. It reads
\begin{equation}
    h({\bf k},k_3) = h_0({\bf k},k_3) + h_1({\bf k},k_3) + \gamma \tau_x \sigma_0,
\end{equation}
where
\begin{equation}
\begin{split}
     h_0(\mathbf{k},k_3) = & \{\epsilon_0 - t\left[ \cos k_1 + \cos k_2 + \cos(k_1 + k_2)\right] \\
     & + \beta \cos k_3 \} \Gamma_1 +\beta \sin k_3 \, \Gamma_3 +
\\ & \xi [\sin k_1 + \sin k_2 - \sin(k_1 + k_2)]\Gamma_3 -\mu \tau_0\sigma_0;
\end{split}
\end{equation}
\begin{equation}
h_1(\mathbf{k},k_3) = \alpha(1-\cos k_3) \big[ \sin k_1\,\Gamma_2 +\sin k_2\,\Gamma_{2,1} -\sin(k_1+k_2)\,\Gamma_{2,2} \big],
\label{eq:bulk_soc}
\end{equation}
the latter being the spin-orbit coupling, while $\gamma \tau_x \sigma_0$ is an inversion-symmetry-breaking term.

The $\Gamma_i$ matrices are products of the Pauli matrices $\sigma_i$ and $\tau_i$, specifically:
\begin{equation}
\Gamma_1 = \tau_z \sigma_0, \quad
\Gamma_2 = \tau_x \sigma_x, \quad
\Gamma_3 = \tau_y \sigma_0,
\end{equation}
and
\begin{equation}
\Gamma_{2,j} = C_3^j \Gamma_2 C_3^{-j},
\qquad
C_3 = \tau_0 \exp\!\left(-i \frac{\pi}{3} \sigma_z \right).
\end{equation}
Here, $\sigma_i$ correspond to the spin degree of freedom, while $\tau_i$ parametrize the orbital degree of freedom. The matrix $C_3$ represents a rotation around the $z$-axis by $2\pi/3$. 

We adopt the convention $k_i = \mathbf{a}_i \cdot \mathbf{k}$, where $\mathbf{a}_i$ are the primitive lattice vectors
$$
\mathbf{a}_1 = (0,1,0), \quad 
\mathbf{a}_2 = \left(\frac{\sqrt{3}}{2}, -\frac{1}{2}, 0\right), \quad 
\mathbf{a}_3 = (0,0,1).
$$
We also explicitly highlight the NN hopping $t$ (set to 1 in the same Ref.).

To observe surface Fermi arcs, we introduce a termination along the $z$-axis by performing an inverse Fourier transform along that direction.
Defining 
\begin{equation}
q(\mathbf{k}) = \sin k_1\,\Gamma_2 +\sin k_2\,\Gamma_{2,1} -\sin(k_1+k_2)\,\Gamma_{2,2},
\end{equation}
the bulk Hamiltonian can be decomposed as
\begin{equation}
h(\mathbf{k},k_3) = h_{\parallel}(\mathbf{k}) + e^{ik_3}V(\mathbf{k}) + e^{-ik_3}V^\dagger(\mathbf{k}),
\label{eq:bulk_layer_decomposition}
\end{equation}
where
\begin{equation}
\begin{split}
h_{\parallel}(\mathbf{k}) ={} & \left\{ \epsilon_0 -t\left[ \cos k_1+\cos k_2+\cos(k_1+k_2) \right] \right\}\Gamma_1
\\
&+ \xi\left[ \sin k_1+\sin k_2-\sin(k_1+k_2) \right]\Gamma_3
\\
&+\alpha q(\mathbf{k}) +\gamma\tau_x\sigma_0 - \mu \tau_0\sigma_0,
\end{split}
\end{equation}
and
$ V(\mathbf{k}) = \frac{\beta}{2}\Gamma_1 + \frac{\beta}{2i}\Gamma_3 - \frac{\alpha}{2}q(\mathbf{k}). $

For a slab containing $N_z+1$ layers, the resulting Hamiltonian has the
block-tridiagonal form
\begin{equation}
H_{\mathrm{slab}}(\mathbf{k})
=
\begin{pmatrix}
h^{\rm cap}_{\parallel} & V & 0 & \cdots \\
V^\dagger & h_{\parallel} & V & \cdots \\
0 & V^\dagger & h_{\parallel} & \cdots \\
\vdots & \vdots & \vdots & \ddots
\end{pmatrix},
\label{eq:slab_hamiltonian}
\end{equation}

where $h^{\rm cap}_{\parallel}({\bf k})$ is the modified top-layer Hamiltonian obtained by changing the in-plane hopping as
\begin{equation}
h_{\parallel}^{\mathrm{cap}}(\mathbf{k}) = h_{\parallel}(\mathbf{k}) - t^\prime \left[ \cos(2k_1)+\cos(2k_2) +\cos\bigl(2(k_1+k_2)\bigr) \right]\Gamma_1 .
\label{eq:cap_hamiltonian}
\end{equation}
We do not consider any change in the interlayer hopping $V$ for the capped layer. This modification is localized at the surface and preserves time-reversal and $C_{3v}$ symmetry, while changing the dispersion of the Fermi arcs. For all the calculation shown in the text, we set $ \epsilon_0=0.4,\, \beta=-0.65,\, \xi=1.2,\, \alpha=-0.25,\, \gamma=-0.25,\, t^\prime=0.6 $ (in units of $t$).

\section{Van Hove singularity}
\label{appendix:vhs}

To highlight the van Hove singularity, we present a 3D rendering of the energy-momentum dispersion of the surface states, i.e. $E(\mathbf k)$, where $\mathbf k = (k_x,k_y)$. Since a surface Hamiltonian for Fermi arcs cannot be defined, the 3D surface is obtained as follows: 
\begin{enumerate}
    \item We compute the surface spectral function $A(\mathbf{k}, \omega)$ on a small square mesh around the Fermi arcs at finite energies $\omega \in [0,0.16]$ (in units of energy).
    \item we extract 
    the points of maximum spectral weight
    for each energy cut. This gives us a 1D contour of the points of highest intensity for each energy slice.
    \item we plot the collected points in three dimensions.
\end{enumerate}

From the plot below we clearly observe a saddle point at roughly $\mathbf k = (0.37,0)$.
\begin{figure}[ht]
    \centering
    \includegraphics[width=0.55\linewidth]{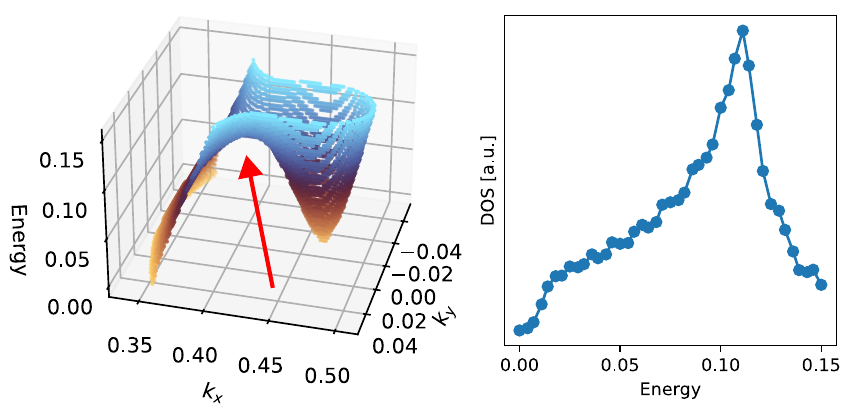}
    \caption{Left: 3D rendering of the surface energy-momentum dispersion relation showing a saddle point (red arrow). The saddle point corresponds to the van Hove singularity. Right: (non-normalized) density of states corresponding to the dispersion shown in the left panel.}
    \label{fig:3d_dispersion}
\end{figure}
To confirm the nature of the van-Hove singularity we also plot the (non-normalized) surface density of state. This quantity is obtained by summing the spectral function over $\mathbf{k}$: $\text{DOS}(\omega) = \sum_{\mathbf k} A(\mathbf k, \omega)$.

We note the similarity between the DOS shown in Fig.~\ref{fig:3d_dispersion} and the surface gap $U\Delta$ in Fig.~\textcolor{blue}{4} of the main text.

\section{Self-consistent surface superconductivity}
\label{appendix:sc}

In this appendix, we provide additional details on the self-consistent procedure.

{Following Ref.~\cite{Trama2025}, superconductivity is introduced by adding to the normal-state Hamiltonian $H_0$ a local attractive interaction,
\begin{equation}
V=-U\sum_{\mathbf r,z,\alpha}
c^{\dagger}_{\mathbf r z\alpha\downarrow}
c^{\dagger}_{\mathbf r z\alpha\uparrow}
c_{\mathbf r z\alpha\uparrow}
c_{\mathbf r z\alpha\downarrow},
\end{equation}
where $U>0$ is the pairing strength and $\alpha$ labels the orbital degree of freedom. We restrict here to an intra-orbital, onsite pairing interaction. Defining
\begin{equation}
\Delta_{z\alpha}
=
\left\langle
c_{\mathbf r z\alpha\downarrow}
c_{\mathbf r z\alpha\uparrow}
\right\rangle
\equiv \Delta_z ,
\end{equation}
the standard mean-field decoupling gives
\begin{equation}
V_{\rm MF}
=
-U\sum_{\mathbf k,z,\alpha}
\left(
\Delta_z\,
c^{\dagger}_{\mathbf k z\alpha\uparrow}
c^{\dagger}_{-\mathbf k z\alpha\downarrow}
+\mathrm{h.c.}
\right)
+
U\sum_{z}|\Delta_z|^2 .
\end{equation}}

{More general local interactions could in principle also include inter-orbital pairing channels. Here we adopt the intra-orbital interaction as the minimal choice: it already reproduces the relevant gap and critical-temperature scales at the correct order of magnitude and is sufficient to capture the surface-enhancement mechanism discussed in this work. In the absence of experimental evidence requiring a dominant inter-orbital pairing channel, we therefore expect this minimal interaction to retain the relevant physics.}

The explicit expressions of $F_B$ and $F_S$ in Eq.~(\textcolor{blue}{4}) of the main text are
\begin{equation}
    F_B = N U \Delta_0^2 \\
- N\frac{k_B T}{2} \int \frac{d^2 \mathbf{k} \, d k_z}{8\pi^3} \sum_i \log \left( 1 + e^{-\beta E_i^B(\mathbf{k}, k_z)} \right),
\end{equation}

\begin{equation}
F_S = 2U \frac{\Delta_0 \Delta_1}{1 - e^{-1/\lambda}} 
+ U \frac{\Delta_1^2}{1 - e^{-2/\lambda}}
- \frac{k_B T}{2} \int \frac{d^2 \mathbf{k}}{4\pi^2} \sum_n \log \left( 1 + e^{-\beta E_n^S(\mathbf{k})} \right),
\label{eq:free_intes}
\end{equation}
$N$ is the number of transverse planes, and $E^B_n(\mathbf{k},k_z)$ and
$E^S_n(\mathbf{k})$ denote the bulk and surface Bogoliubov de Gennes eigenvalues
\begin{align}
&E_n^B(\mathbf{k},k_z) = \pm\sqrt{\left(\varepsilon^B_n(\mathbf{k},k_z)\right)^2+U^2\Delta_0^2}, \\
&E_n^S(\mathbf{k}) = \pm\sqrt{\left(\varepsilon^S_n(\mathbf{k})\right)^2+U^2(\Delta_0+\Delta_1 f_{n,\lambda}(\mathbf{k}))^2}.
\label{eq:BdgSurf}
\end{align}
Here, $\varepsilon^B_n(\mathbf{k},k_z)$ and $\varepsilon^S_n(\mathbf{k})$ are the bulk and surface eigenvalues of the normal-state Hamiltonian. 
The factor $f_{n,\lambda}(\mathbf{k})=\sum_{\alpha,z} |c^n_{\alpha z}|e^{-z/\lambda}$ represents the weight of each state contributing to surface superconductivity, where $c^n_{\alpha z}$ are the coefficients of the wavefunctions in the chosen basis $\ket{\psi^S_n}=\sum_{\alpha z}c^n_{\alpha z}\ket{\varphi_{\alpha z}}$.
Note how Eq.~\eqref{eq:BdgSurf} helps us \textit{rigorously} defining the projected surface order parameter (gap): $g_n(\mathbf k) = U(\Delta_0 + \Delta_1 f_{n,\lambda}(\mathbf k))$. This object acquires a momentum dependence despite both $\Delta_1$ and $\Delta_0$ being constant because of the surface state projection in $f_{n,\lambda}$.

In the thermodynamic limit ($N\to \infty$), $F_S/F_B\to 0$; {therefore, the self-consistent value of the bulk gap can be obtained by minimizing only $F_B$ with respect to $\Delta_0$, as it is independent of the surface properties as
\begin{equation}
    \frac{\partial F_B(\Delta_0)}{\partial \Delta_0}=0.
\end{equation}}

For a real material, the value of $U$ depends on the specific microscopic coupling that leads to attractive pairing (e.g., electron-phonon or electron-electron). However, within our framework, it can be treated as a free parameter. We therefore determine the value of $U$ that yields a desired bulk gap $U\Delta_0$ at $T=0$~K~\footnote{For numerical reasons, we set the temperature to $T=10^{-8}t$ ($t$ being the nearest neighbor hopping) .}. {Once $U$ is fixed, we minimize $F_S$ with respect to $\Delta_1$ and $\lambda$, interpreting $F_S$ as the surface contribution to the free energy as
\begin{equation}
    \frac{\partial F_S(\Delta_0,\Delta_1,\lambda)}{\partial \Delta_1}=0,
    \qquad
    \frac{\partial F_S(\Delta_0,\Delta_1,\lambda)}{\partial \lambda}=0.
    \label{eq:min_sur}
\end{equation}}

In Eq.~\eqref{eq:free_intes}, the value of $\Delta_0$ plays a crucial role in determining the other two parameters. 
Note that the superconducting gap at the surface is enhanced only within a
certain window of $U$.  For large values of $U\Delta_0$, the linear term
$U\Delta_0\Delta_1$ makes $F_S$ positive, suppressing any enhancement of the
surface superconductivity. That is, when the system has a strong tendency
towards a superconducting instability, the highest energy gain comes from
realizing a bulk superconductor.

We therefore seek a regime where $\Delta_0$ is small but finite, allowing
$\Delta_1$ to be sizable, while the bulk remains metallic.
In this context, Weyl semimetals provide an ideal platform: due to the low
density of states near the Weyl points at the Fermi level, even relatively large
$U$ yields only a small bulk gap at finite temperature ($U\Delta_0 <
10^{-4}$~eV). At the same time, because of the topological surface states,
$\Delta_1$ can be sizable.
We stress that the absence bulk superconductivity here is not a drawback but a
key feature: if the system becomes fully superconducting, the tunability
associated with the surface is lost.

In Fig.~\ref{fig:UD} we show the behavior of $U\Delta_0$ and $U\Delta_1$ as functions of $U$. A crossover between bulk and surface contributions occurs around $U \sim 1.44$.
\begin{figure}[t]
    \centering
    \includegraphics[width=0.45\linewidth]{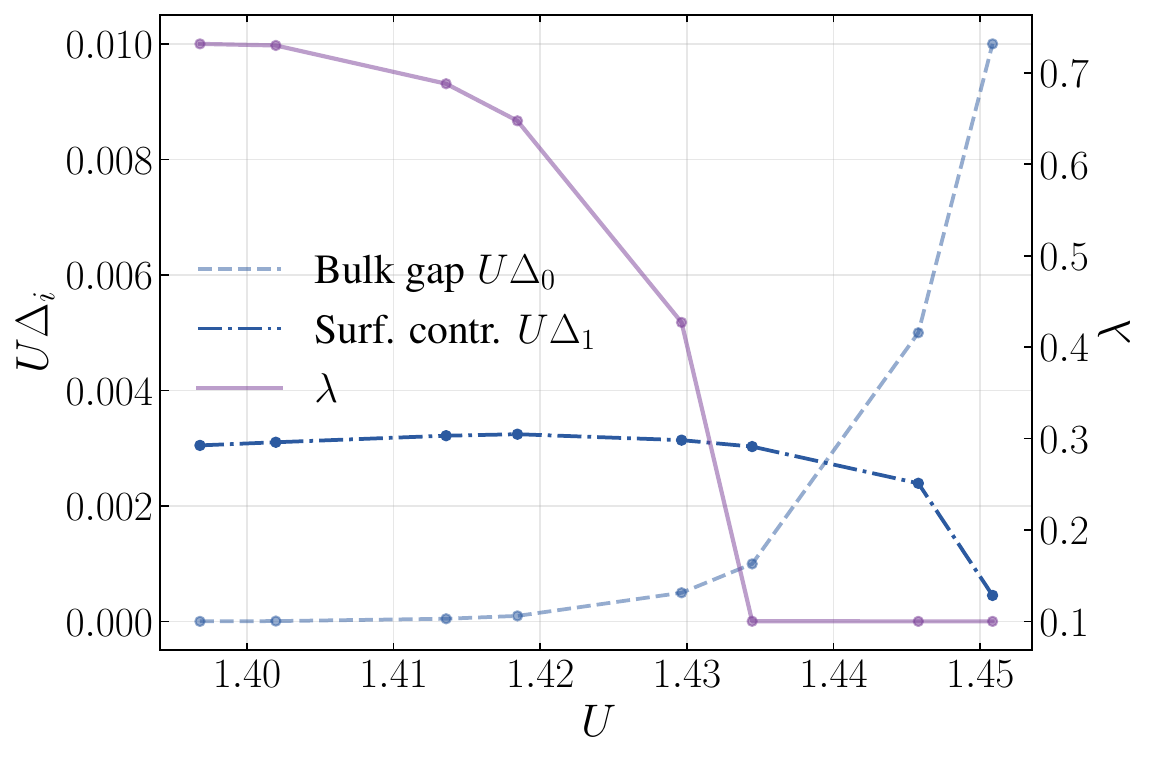}
    \caption{Gap contributions $U\Delta_0$ and $U\Delta_1$, and the surface penetration length $\lambda$, as functions of the self-consistent superconducting strength $U$.}
    \label{fig:UD}
\end{figure}
\begin{figure}[t!!]
    \centering
    \includegraphics[width=0.34\linewidth]{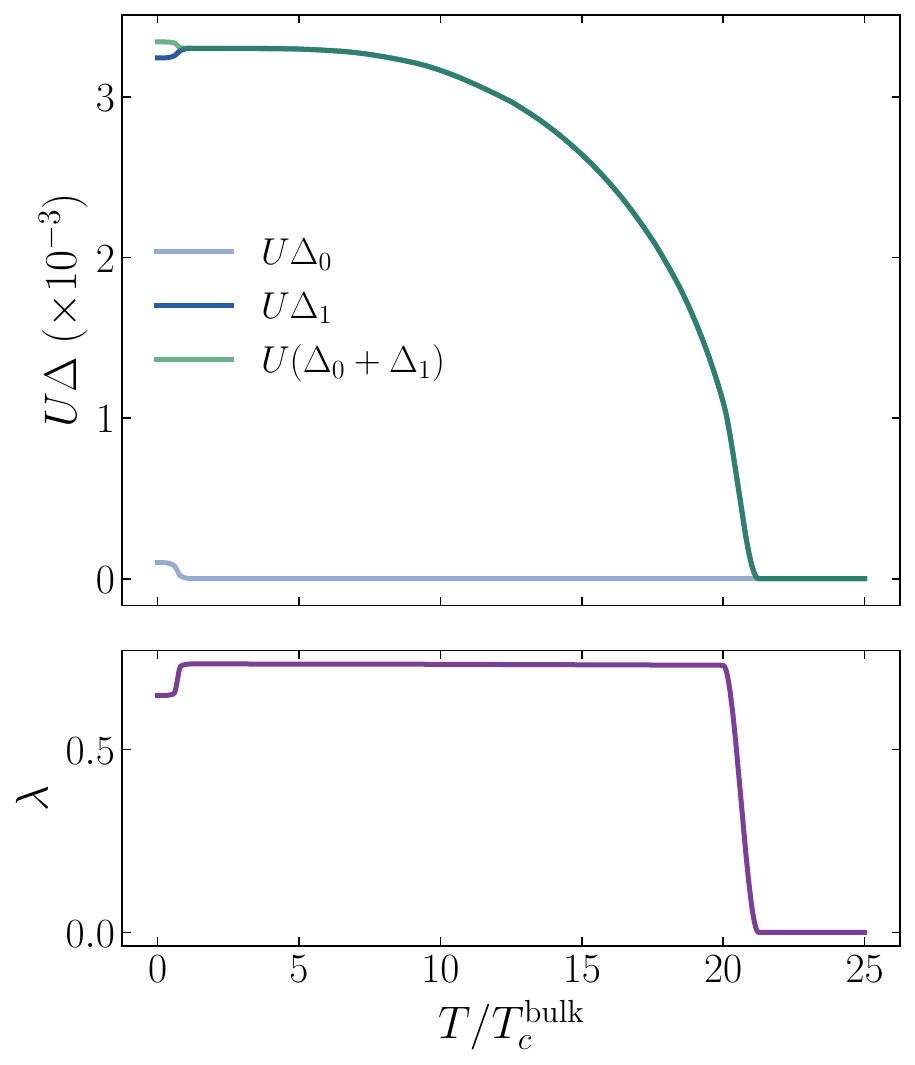}
    \caption{(Upper panel) Gap contributions $U\Delta_0$ and $U\Delta_1$, and their sum and (lower panel) surface penetration length $\lambda$, as functions of temperature $T$ expressed in unit of the $T_c^{\mathrm{bulk}}$ computed for $\mu$ at the VHS.}
    \label{fig:U_vs_T}
\end{figure}
The remaining discussion is devoted to the critical temperature analysis of the system.
In Fig~\ref{fig:U_vs_T} we show the temperature dependence of the superconducting gap opened in the bulk and at the surface. The $x$-axis is given in units of the bulk critical temperature $T_c^{\text{bulk}}= 10^{-5}t$ ($t$ being the NN hoppings). The surface $T_c$ is 21 times higher. To match the experimental $T_c^{\text{bulk}} \sim 0.6$~K in PtBi$_2$ we should consider $t \sim 0.75$~eV, yielding $T_c^{\text{surf}} \sim 13$~K, in agreement with Scanning Tunneling Microscope and ARPES data~\cite{Schimmel2024,Kuibarov2024}.
Note that there exist only one transition temperature $T_c$ and the division between $T_c^{\text{bulk}}$ and $T_c^{\text{surf}}$ is mainly for bookkeeping purposes. That is, when superconductivity is exponentially localized at the surface, we talk about surface critical temperature. On the other hand, when the order parameter develops an homogeneous component in the system we refer about $T_c^{\text{bulk}}$.

\section{Spin texture of the arcs and effective gap symmetry}
\label{app:spin_gap_symmetry}

The local attractive interaction considered in this work produces an $s$-wave mean-field order parameter transforming according to the fully symmetric $A_1$ representation of $C_{3v}$. Nevertheless, the effective gap opened on the Fermi arcs is not momentum independent. Indeed, the spin-orbit coupling and the broken inversion symmetry lock the spinor wave function of the arc state to its momentum. Consequently, projecting the local $s$-wave solution onto the nondegenerate surface states generates both singlet- and triplet-like
components in the effective gap matrix. Together with the inhomogeneity, these ingredients generate a gap amplitude that changes along the Fermi surface.

Eq.~\eqref{eq:BdgSurf} shows this explicitly through the factor $f_{n,\lambda}(\mathbf{k})$, which accounts for the overlap between the penetration profile of the Fermi arcs and the order parameter. Fig.~\ref{fig:effective_gap_arc}(a) shows the gap $g({\bf k})=U[\Delta_0 + \Delta_1f_\lambda({\bf k})]$ opened on one of the Fermi arcs at the VHS (the band index $n$ is determined by the band crossing the Fermi level at each $\mathbf{k}$ point along the Fermi arc).

The spinorial decomposition of the gap matrix $\widetilde{\Delta}$~\cite{Vocaturo2024} emerges clearly in the density matrix representation; our subsequent discussion therefore reformulates the arguments of Ref.~\cite{Vocaturo2024} and generalizes them to include the $\bf k$-dependent penetration of the Fermi arcs into the material.
The fully symmetric $A_1$ pairing
sector can be described in terms of three matrix-valued lattice
harmonics: the singlet $s$-wave structure $d_0(\mathbf{k})$, the
in-plane Rashba-like triplet $p$-wave structure
$d_{xy}(\mathbf{k})$, and the Ising-like triplet $f$-wave structure $d_z(\mathbf{k})$, which take the form
\begin{align}
d_0(\mathbf{k}) &= \sigma_0,
\\
d_{xy}(\mathbf{k})
&=
\sin(k_1+2k_2)\,\sigma_y
-\sin(2k_1+k_2)\,
 C_3\sigma_y C_3^\dagger
\nonumber\\
&\quad
+\sin(k_1-k_2)\,
 C_3^2\sigma_y
( C_3^\dagger)^2,
\\
d_z(\mathbf{k})
&=
\Big[
\sin(k_1+2k_2)
-\sin(2k_1+k_2)
+\sin(k_1-k_2)
\Big]\sigma_z,
\label{eq:surface_gap_harmonics}
\end{align}
where
$ C_3
=
e^{-i\pi\sigma_z/3}
$.

A normalized surface eigenstate can be written as
$
\ket{\psi_{n\mathbf{k}}}
=
\sum_{\alpha,z,\sigma}
c^n_{\alpha z\sigma}(\mathbf{k})
\ket{\alpha,z,\sigma},
$
where $\alpha$, $z$, and $\sigma$ denote orbital, layer, and spin
indices, respectively, with the requirement that $|c^n_{\alpha z\sigma}(\mathbf{k})|^2$ has a significant peak over the first few layers (in the manuscript we chose $\sum_{z=0}^5 |c^n_{\alpha z\sigma}(\mathbf{k})|^2\geq0.5$). Tracing over the orbital and layer degrees of
freedom gives the reduced spin-density matrix
\begin{equation}
\left[\rho^{\mathrm{spin}}_{n\mathbf{k}}\right]_{\sigma\sigma'}
=
\sum_{\alpha,z}
c^n_{\alpha z\sigma}(\mathbf{k})
\left[c^n_{\alpha z\sigma'}(\mathbf{k})\right]^*
=
\frac{1}{2}
\left[
\sigma_0+\mathbf S_n(\mathbf{k})\cdot\boldsymbol{\sigma}
\right],
\label{eq:reduced_spin_density}
\end{equation}
where $\mathbf S_n(\mathbf{k})$ is the spin polarization of the arc
state.

To characterize the effective gap on the Fermi surface, we project the gap operator at each wave vector $\mathbf{k}$ onto the corresponding surface state: the gap operator at the wavevector $\bf k$ results from a projection onto the eigenstate:
\begin{equation}
\mathcal M_n(\mathbf{k})
=
g_n(\mathbf{k})\rho^{\mathrm{spin}}_{n\mathbf{k}}
=
\eta_s(\mathbf{k})d_0(\mathbf{k})
+
\eta_p(\mathbf{k})d_{xy}(\mathbf{k})
+
\eta_f(\mathbf{k})d_z(\mathbf{k}).
\label{eq:harmonics}
\end{equation}
The coefficients $\eta_s$, $\eta_p$, and $\eta_f$ quantify the
$s$-, $p$-, and $f$-like content of the effective matrix. In the
conventional singlet--triplet notation, the corresponding gap matrix is
\begin{equation}
\widetilde{\Delta}_n(\mathbf{k})
=
\mathcal M_n(\mathbf{k})U_T,
\qquad
U_T=i\sigma_y.
\label{eq:effective_gap_matrix_referee}
\end{equation}

{Here, $M_n(\mathbf{k})$ and $U_T$ act on the same spin space, so Eq.~(25) denotes an ordinary $2\times2$ matrix multiplication and not a tensor product between independent Pauli spaces.}

Note that comparing Eq.~\eqref{eq:reduced_spin_density} to Eq.~\eqref{eq:harmonics} one sees clearly how the spin texture dictates the weights of the relative channel. This implies, for example, that in our model there can not be a sizable $f$-wave part since the spin along the arc lie all in the $xy$ plane (see Fig.~\ref{fig:effective_gap_arc}(c)).
In Fig.~\ref{fig:effective_gap_arc}(b), we plot
$\eta_p(\mathbf{k})$ along the selected Fermi arc.
This quantity represents the $p$-like part of the effective gap matrix. 
We emphasize that $\eta_pd_{xy}$ is induced by projecting the mean-field solution for a fully symmetric local interaction
onto the spin-polarized Fermi-arc
states and it is not an independent pairing channel.

\begin{figure}
    \centering
    \includegraphics[width=0.99\textwidth]{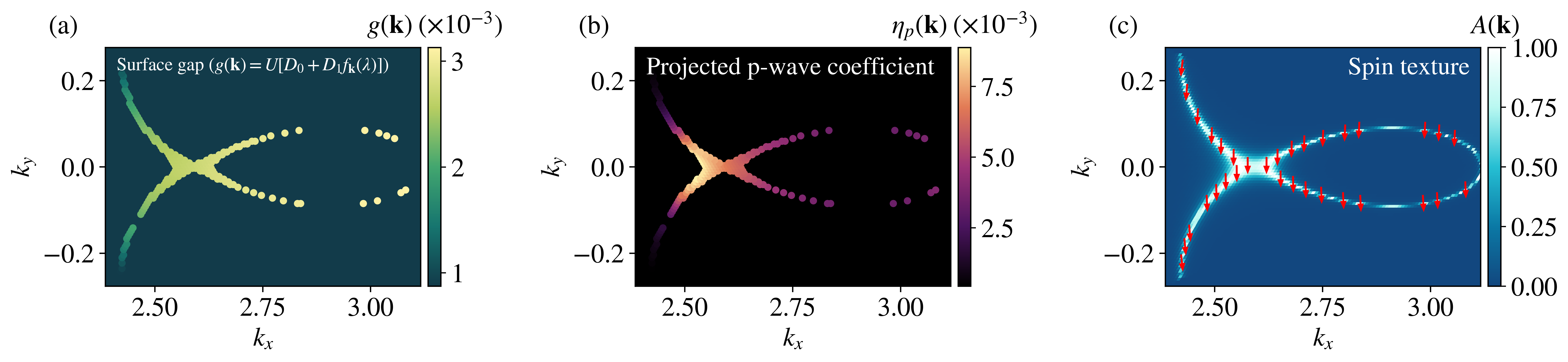}
    \caption{
    Effective superconducting structure of the surface arcs at
    $\mu=0.11$, corresponding to the surface van Hove singularity.
    (a) Scalar surface gap
    $g(\mathbf{k})=U[\Delta_0+\Delta_1f_{\lambda}(\mathbf{k})]$.
    (b) Coefficient $\eta_p(\mathbf{k})$ of the induced
    $p$-like contribution to the effective gap matrix.
    (c) Normalized surface spectral function; the arrows indicate the
    direction of the in-plane spin polarization.
    }
    \label{fig:effective_gap_arc}
\end{figure}

\twocolumngrid
\bibliography{bib}

\end{document}